# Concepts of super-valley electron and twist induced quantum super-valley Hall effect

Yu-Hao Shen[1,†], Jun-Ding Zheng[1,†], Chun-Gang Duan[1,2*]

[1] *Key Laboratory of Polar Materials and Devices, Ministry of Education, Department of Electronics, East China Normal University, Shanghai, 200241, China*

[2]*Collaborative Innovation Center of Extreme Optics, Shanxi University, Taiyuan, Shanxi 030006, China*

† Equally contributed authors

* **E-mail:** cgduan@clpm.ecnu.edu.cn

## Abstract

Collective motions of electrons in solids are often conveniently described as the movements of quasiparticles. Here we show that these quasiparticles can be hierarchical. Examples are valley electrons, which move in hyperorbits within a honeycomb lattice and forms a valley pseudospin, or the self-rotation of the wave-packet. We demonstrate that twist can induce higher level motions of valley electrons around the moiré superlattice of bilayer systems. Such larger scale collective movement of the valley electron, can be regarded as the self-rotation (*spin*) of a higher-level quasiparticle, or what we call super-valley electron. This quasiparticle, in principle, may have mesoscopic size as the moiré supercell can be very large. It could result in fascinating properties like topological and chiral transport, superfluid, *etc*., even though these properties are absent in the pristine untwisted system. Using twisted antiferromagnetically coupled bilayer with honeycomb lattice as example, we find that there forms a Haldane-like superlattice with periodically staggered magnetic flux and the system could demonstrate quantum super-valley Hall effect. Further analyses reveal that the super-valley electron possesses opposite chirality when projected onto the top and bottom layer, and can be described as two components (magnetic monopoles) of Dirac fermion entangled in *real-space*, or a giant electron. Our theory opens a new way to understand the collective motions of electrons in solid.

Crystal environment provides excellent platform to modify the behaviors of electrons. The so formed quasiparticles, or the collective motions of electrons, could demonstrate fascinating properties like superconducting, spin liquid, quantum Hall effect, *etc* [1-5]. Moreover, by introducing the topological properties, we can simulate the elementary particles, which generally require extremely high energy to be studied experimentally. Successful examples include Weyl fermion [6, 7], Majorana fermion [8, 9], and even Higgs boson [10]. Such simulation is possible relying on the fact that the wavefunctions of these quasiparticles in the crystals or solids satisfy essentially the same Schrödinger or Dirac equations as those elementary particles do.

The analogy between the quasiparticles and elementary particles so far, however, currently are only found in momentum space. This puts a strict limitation on simulating the elementary particles, which in fact are existing in real-space. By the way, to imagine the movement of the quasiparticle in momentum space is also not straightforward. Moreover, if we can observe these elementary particle-like quasi particles in real-space, we could observe their both stationary and dynamic behaviors which may be even more exotic.

In this study, we propose that it is possible to sketch the quasiparticles formed by the collective motions of electrons in real-space. And, amazingly, we find that these quasiparticles indeed have hierarchy, i.e., the collective motion of the lower-level quasiparticle could form a higher-level quasiparticle. At different level, the quasiparticles may have quite different behaviors, i.e., they may obey different wave equations, corresponding to different elementary particles. More interestingly, through the upgrade of the quasiparticle, it will equip more fascinating properties which are absent in the lower-level quasiparticle. This well interprets the philosophy of *More is different* [11], and will be demonstrated in following section.

To begin the story, we first introduce the concept of valley electron as a quasiparticle. It is well known that, in graphene-like systems, there is an intrinsic magnetic moment associated to the valley electron, which is called the valley pseudospin in momentum space [12-15]. Indeed, this valley pseudospin is formed by the circling movement of valley electrons in the hyperorbit of a honeycomb lattice, or the self-rotation from the corresponding wave-packet [16]. Note that the circular motion of the valley electron will generate a magnetic flux, which has a minimum quantized value *h/2e* and corresponds to a spin number $\tau=\frac{1}{2}$. In addition, the spinor wavefunction of the valley electron satisfies the two-dimensional Dirac equation, rendering the valley spin an intrinsic property of the valley electron, just its two components represent $\mathbf{K}_+$ (valley spin-up) and $\mathbf{K}_-$ (valley spin-down) states. The valley electron has charge and spin as common electron does, but it has a much larger size, since it covers the whole honeycomb lattice. Consequently, the valley electron with new understanding can be regarded as a large size fermion. In this sense, we can see that the so-called valley Hall effect is essentially spin Hall effect, just this spin is valley spin which belongs to the valley electron. It is very convenient to use this picture to explain the phenomenon related to valley physics, e.g., the valley index is good quantum number due to the strongly suppressed intervalley scattering, which now can be interpreted as the weak general spin-orbit coupling (SOC) i.e., *valley spin*-orbit coupling for the valley electron system, therefore $K_+$ and $K_-$ electrons are hard to be mixed. We should mention that this concept is not a trivial re-interpretation of the valley electron. It not only provides a real-space picture of the valley electron, but can be extended to further degree of the electron motions, e.g., electrons in a moiré superlattice [17, 18]. As shown in Fig. 1, there will form higher level valley electron with enlarged size. Finally, as we will show later, it may help us to understand the inner structure of elementary particles or how the particles are packed by more fundamental particles, through phase connection of wavefunctions.

We further point out that the wavefunction for individual valley electron of the graphene-like system preserves U(1) symmetry, as additional phase factor $e^{i\phi}$ does not change the dynamics of the valley electron since the wavefunction is enclosed in the honeycomb lattice. We can also say that this U(1) gauge transformation describes the planar phase connection of electron wavefunctions within the atomic sites of the honeycomb lattice, which gives rise to the Dirac wave-packet of valley electron [16, 19]. In the case when the two opposite valley electrons are decoupled, i.e., the *valley spin*-orbit coupling or inter-valley scattering is negligible, the system indeed possesses U(1)×U(1) symmetry. However, if the *valley spin*-orbit coupling becomes significant, two valleys with opposite chirality can be phase connected and valley states now should be described as spinor field (SU(2) gauge field), i.e., the valley system now satisfy SU(2) symmetry. This is very interesting since it indicates that the system may demonstrate quantum spin Hall (QSH) effect, as predicted by Kane and Mele [20, 21].

Here we propose that twist effect could introduce effective *valley spin*-orbit coupling interaction, and may connect the wavefunctions of the valley electrons to form a much larger wave-packet. We call this wave-packet *super-valley* electron, as it is formed by the collective motion of valley electrons. This clearly indicates the existence of the hierarchy of the quasiparticles, which is exactly the same as the architecture of the elementary particle systems. Moreover, following our strategy as described below, the elementary particle-like quasiparticles can be shown in *real-space*, and confirms the relationship between topology and the composition of particles.

Let us start with the relativistic Dirac equation, i.e.

$$i\hbar\gamma^{\mu}\partial_{\mu}\Psi=mc\Psi \qquad (1)$$

where $\gamma^\mu$ is the 4 × 4 Gamma matrices, $m$ and $c$ are the electron mass and the velocity of the light, respectively. It is well known that when $m$ vanishes, the solution of this equation $\Psi$ is actually a bispinor which consists of two Weyl spinors with opposite chirality, i.e., $\Psi=(\varphi_R,\varphi_L)$ with $\varphi_{R(L)}$ being the right (left)-handed solution of the Weyl equation which describes massless fermion,

$$i\hbar\sigma^\mu\partial_\mu\varphi=0 \qquad (2)$$

with $\sigma^\mu$ being the Pauli matrices (quaternion).

Note that although the above equations are initially defined in the four-dimensional space-time, it can be extended to more general four-dimensional space, e.g., a Hilbert space spanned by the conduction and valence bands with spin up and down indices (a two-band model). Interestingly, the valley system, which has energy maximum in the valence band and energy minimum in the conduction band at the same $k$-point, well satisfies the Dirac-like relativistic electron equation. When the gap between the conduction and valence band closes, the solutions of this Dirac-like equation, i.e., electron and hole (anti-electron or positron) become energy degenerate and massless, same as the electron in the Dirac cone of the graphene system. We should point out that the origin of the mass of the Dirac electron is the interaction between two massless Weyl fermions with opposite chirality which breaks the chiral symmetry [22], similar as the Higgs mechanism [23].

Inspired by this spirit, we propose that the same strategy can be extended to the real-space, e.g., a bilayer valley system with honeycomb lattice. Different from previous studies, we require the system to be ferromagnetic intra-layer coupling and antiferromagnetic inter-layer coupling, so that the two degenerate spin channels $s$ can be spatially separated. When we further introduce the SOC effect, the individual layer both becomes ferrovalley [24] and there emerges the valley spin $\tau$ from two layers that are also antiferromagnetically coupled [25, 26]. Haldane [27] and Kane-Mele [20] previously also demonstrate that two-component spinor constructed on A and B sublattices in graphene system, which is also defined in real-space, is useful to explain quantum (spin) Hall effect. Indeed, our model system is a two-layer counterpart of the Haldane and Kane-Mele model (Fig. 2a), just the hopping between A and B sublattices is substituted by the top and bottom layer tunneling (Fig. 2b). We further point out that the essence of realizing the QSH effect is to form a parity(chirality)-anomaly particle with charge $e$ and 1/2 spin which can be described by Eq. (1) and circumvents the *fermion doubling* problem in solids [27].

Previous studies indicate that twist will induce a layer dependent gauge vector potential $A$ in bilayer graphene-like systems [28, 29]. A nontrivial gauge connection $A$ can be chosen as a skyrmion-like texture enclosed in the honeycomb lattice [30, 31]. We further point out that such induced connection $A$ could result in the mixture of different valley spin $\tau$, i. e., *valley spin*-orbit coupling. To be specific, it is demonstrated that $A$ can be proportional to $\tau_i\times\tau_j$ with moiré period $a_m$ in real-space, where $i$ and $j$ are nearest neighbor sites in an enlarged triangular lattice (see Appendix). Due to stacking variation, in each individual layer of such bilayer system, $\tau$ becomes canted, and its distribution should be continuous at boundary of the moiré superlattice. Under $C_3$ symmetry constraint the out of plane component of $A$ in

one layer satisfies:

$$A_z(\mathbf{r})=A_0 \sum_{\alpha=1,2,3} \sin(\mathbf{G}_\alpha^\parallel \cdot \hat{z}\times\mathbf{r}) \qquad (3)$$

Here, $\mathbf{G}_1^\parallel=\frac{4\pi}{\sqrt{3}a_m}\hat{y}$, $\mathbf{G}_2^\parallel$ and $\mathbf{G}_3^\parallel$ are respectively obtained by counterclockwise rotation of $\mathbf{G}_1^\parallel$ with $2\pi/3$ and $4\pi/3$. $A_0$ characterize the amplitude of $A_z$. And $\mathbf{r}$ is the position vector measured from center. For the other layer we have $A_z \to -A_z$. For a gauge connection $\mathbf{A}$ which is exactly enclosed, we choose the in-plane components as [30, 31]:

$$A_x(\mathbf{r})=A_1[1+\cos(\mathbf{G}_2^\parallel \cdot \hat{z}\times\mathbf{r})+\cos(\mathbf{G}_3^\parallel \cdot \hat{z}\times\mathbf{r})] \qquad (4a)$$

$$A_y(\mathbf{r})=A_1[-\sin(\mathbf{G}_2^\parallel \cdot \hat{z}\times\mathbf{r})+\sin(\mathbf{G}_3^\parallel \cdot \hat{z}\times\mathbf{r})] \qquad (4b)$$

where $A_1$ characterize the in-plane amplitude. Note that $C_3$ symmetry requires $\mathbf{A}$ remains unchanged under the substitution $\mathbf{G}_1^\parallel \to \mathbf{G}_2^\parallel$, $\mathbf{G}_2^\parallel \to \mathbf{G}_3^\parallel$ and $\mathbf{G}_3^\parallel \to \mathbf{G}_1^\parallel$ and in turn. Then, the entire real-space pattern of $\mathbf{A}$ for one layer is shown in Fig. 3a. We can numerically prove that the topological winding number [32]:

$$N_w=\frac{1}{4\pi}\iint \frac{\mathbf{A}}{|\mathbf{A}|^3}\cdot(\partial_x\mathbf{A}\times\partial_y\mathbf{A})=\pm 1 \qquad (5)$$

for top $(+1)$ and bottom layer $(-1)$, respectively per moiré cell (denoted by the hexagon) through the integral of a real-space Berry curvature $\Omega_{\mathbf{r}}^z=\frac{1}{2}\frac{\mathbf{A}}{|\mathbf{A}|^3}\cdot(\partial_x\mathbf{A}\times\partial_y\mathbf{A})$ (shown in Fig. 3b). Therefore, we find the quantized Berry flux $\pm 2\pi$ per moiré cell [30, 31].

Note that this nontrivial gauge connection $\mathbf{A}$ is layer dependent, and it is indeed one component of a non-Abelian gauge filed $A^\mu$ which is defined in a two-dimensional space spanned by sublattice indices A and B. Then a $2\times 2$ interlayer tunneling matrix can be expressed as $U=u_0\exp(ie\mathbf{A}\cdot\boldsymbol{\sigma}/u_0)$ [33], where $u_0$ describes the interlayer tunneling for AB stacking of untwisted case and Pauli matrix $\boldsymbol{\sigma}$ defines the sublattice indices. Under expansion of $U$, we can express the twisted bilayer Hamiltonian for valley electron with momentum $\mathbf{p}$ as

$$H_{bi}=\begin{pmatrix} \mathbf{p}\cdot\boldsymbol{\sigma} & u_0-ie\mathbf{A}\cdot\boldsymbol{\sigma} \\ u_0+ie\mathbf{A}\cdot\boldsymbol{\sigma} & \mathbf{p}\cdot\boldsymbol{\sigma} \end{pmatrix} \qquad (6)$$

Taking four-component spinors constructed on the top $(t)$ and bottom $(b)$ layer $\Psi(t,b)$ as the basis [33], we can transform Eq. (6) into

$$H_{eff}=\begin{pmatrix} \boldsymbol{\pi}_+\cdot\boldsymbol{\sigma} & -iu_0 \\ iu_0 & \boldsymbol{\pi}_-\cdot\boldsymbol{\sigma} \end{pmatrix} \qquad (7)$$

where $\boldsymbol{\pi}_\pm=\mathbf{p}\mp e\mathbf{A}$ [19, 34]. Now it is clear that the solution of Eq. (7) actually corresponds to a Dirac bispinor which consists of two Weyl spinors with opposite chirality combined with their entanglement. The interesting part is that the diagonal terms $\boldsymbol{\pi}_\pm\cdot\boldsymbol{\sigma}$ have exactly opposite sublattice polarizations [33]. Consequently, on the rotation origin (AB stacking regions) they can be projected onto each layer and there give rise to their entanglement. Now we can *see* this Dirac fermion and its components (Weyl fermion or the Dirac monopole) in *real-space*.

The real $u_0$ of the off-diagonal term can describe an effective Dirac mass (multiplied with $\gamma^5$ matrix in Dirac spinor basis of Eq. (6)) due to interactions between massless fermion quasiparticles, which can be realized by chiral anomaly. It means that there forms a parity(chirality)-anomaly particle to circumvent the *fermion doubling* problem. Its real space correspondence can be understood by the lower symmetry of the valley spin $\tau$ structure which is absent of inversion center when mapped from top layer to bottom one. It is also a noncolinear configuration between antiferromagnetically coupled valley spin $\tau$ of two layers. We attribute such symmetry breaking comes from many-body correlation problem of a $SU(4)$ model [35], where a $SU(2)_L \times SU(2)_R \times U(1)_{sv}$ symmetry can be obtained (See Appendix) and it indeed belongs to our single-particle model in Eq. (7). The super-valley $U(1)_{sv}$ symmetry here describes the self-rotation of the super-wavepacket which is the result of the entanglement of the two valley wavepacket with opposite chirality in the moiré cell, and is indeed a massive Dirac electron satisfying Eq. (1).

The above topological nontrivial phase can be interpreted in another way. Actually, as can be seen from Fig. 3, the distribution of $A$ gives rise to the staggered phase (magnetic flux) with moiré period [30, 31] which exactly satisfies Haldane model [27] with a super-honeycomb-lattice. This is another reason we call the corresponding single-particle as super-valley electron, as it describes hyperorbit rotation within a super-honeycomb lattice. The behavior of super-valley electron then can be well explained by Haldane model. In this sense, the condition to approach nontrivial topological phase of Haldane lattice is to modify its model parameters through twist angle until the phase connection $A$ is enclosed in the entire moiré superlattice, in other words there connects the wavefunctions phase between AA and BB regions. In such a super-honeycomb-lattice, the on-site energy of the single-particle in Haldane model for 'AA-site' or 'BB site' reduces as twisting from the pristine AB stacked system (it is indeed zero on-site energy for 'magic angle' system [36]). Therefore, according to the phase diagram of Haldane model, the quantized anomalous Hall conductance for each layer can always be found in such Haldane-like superlattice [37]. The bilayer then becomes QSH system under this consideration, and it is indeed a macroscopic quantum phenomenon since the super-valley electron can be very large. As now the electrons are indeed super-valley electrons, we can call this QSH effect as quantum super-valley Hall effect. And the super-valley spin can be detected using the circularly polarized light or light with orbital angular momentum [38].

From the above analysis, we can see that the introduction of twist brings rich novel physics to the two-dimensional magnetic bilayer system. Twist will destroy the translational symmetry of the pristine lattice, and forms a moiré superlattice with patterned lattice potential, which could satisfy various models, such as Haldane-like or Kane-Mele-like model, in a mesoscopic scale. The resulting *real-space* quasiparticles, i.e., collective motions of electrons in a much larger spatial scale than that of the pristine system, thus can have various forms of fermion (Weyl fermions, Dirac fermions or axions, *etc*.). Surprisingly, we find that there exists hierarchy in these quasiparticles, e.g., common electron, valley electron and super-valley electron as we proposed here. This could be attributed to the hierarchy of the circular motion of the quasiparticles. What more interesting is that, the quasiparticle with higher rank demonstrates more fundamental properties of electrons, like magnification effect. Therefore, with the variation of the crystal potential, we could *create* various elementary particles in a crystal *universe*, which may help us to understand how our world works.

**Acknowledgements**

This work was supported by the National Key Research and Development Program of China (2017YFA0303403), Shanghai Science and Technology Innovation Action Plan (No. 19JC1416700), the NSF of China No. 11774092), ECNU Multifunctional Platform for Innovation.

**FIG. 1** (a) Different stacking regions in a moiré superlattice. (b) Sketch of the formation of supervalley electron with spin ($\tau_S$), which is indeed a higher level wave-packet constructed by valley electrons ($\tau$) in twisted honeycomb lattice.

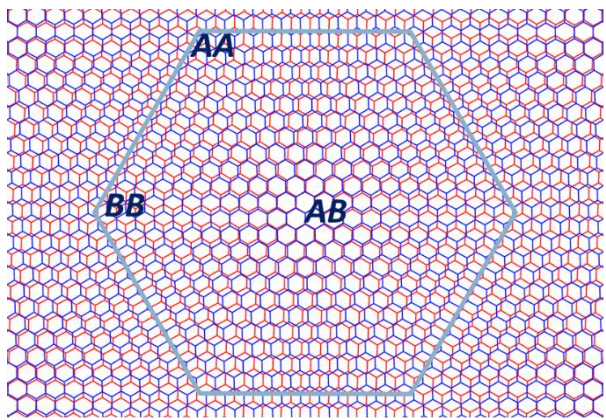
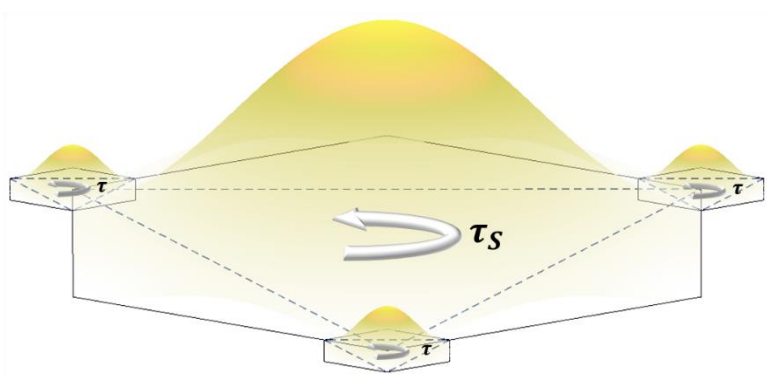

**FIG. 2** A intuitive view of the antiferromagnetically coupled bilayer: the hopping between A and B sublattices is substituted by the tunneling between layer A and B.

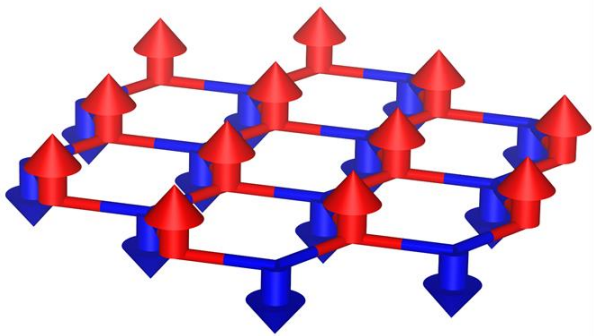
**Sublattice A, B**

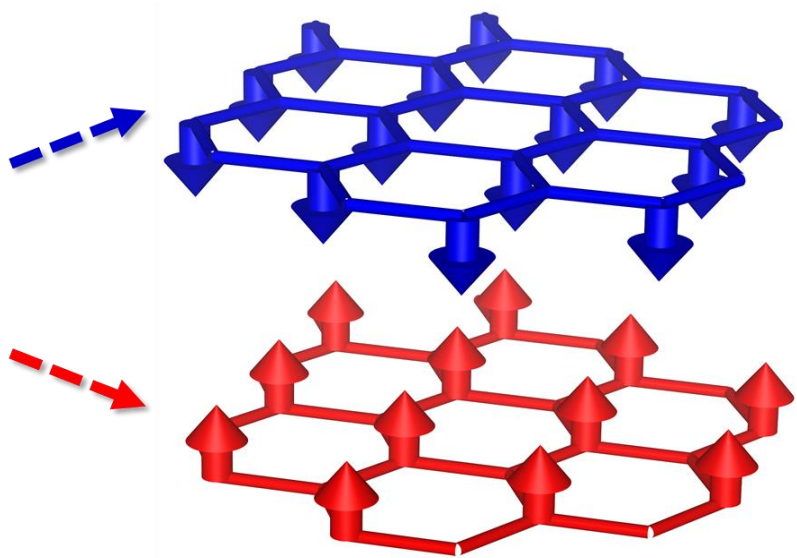
**Layer A, B**

**FIG. 3** (a) Normalized distribution pattern of vector potential $A$ in real-space, here we choose $A_0=A_1$. The in-plane and out-of-plane components are shown in red arrows and color map, respectively. (b) The distribution of real-space Berry curvature $\Omega_r^z$ in moiré superlattice denoted by the hexagon.

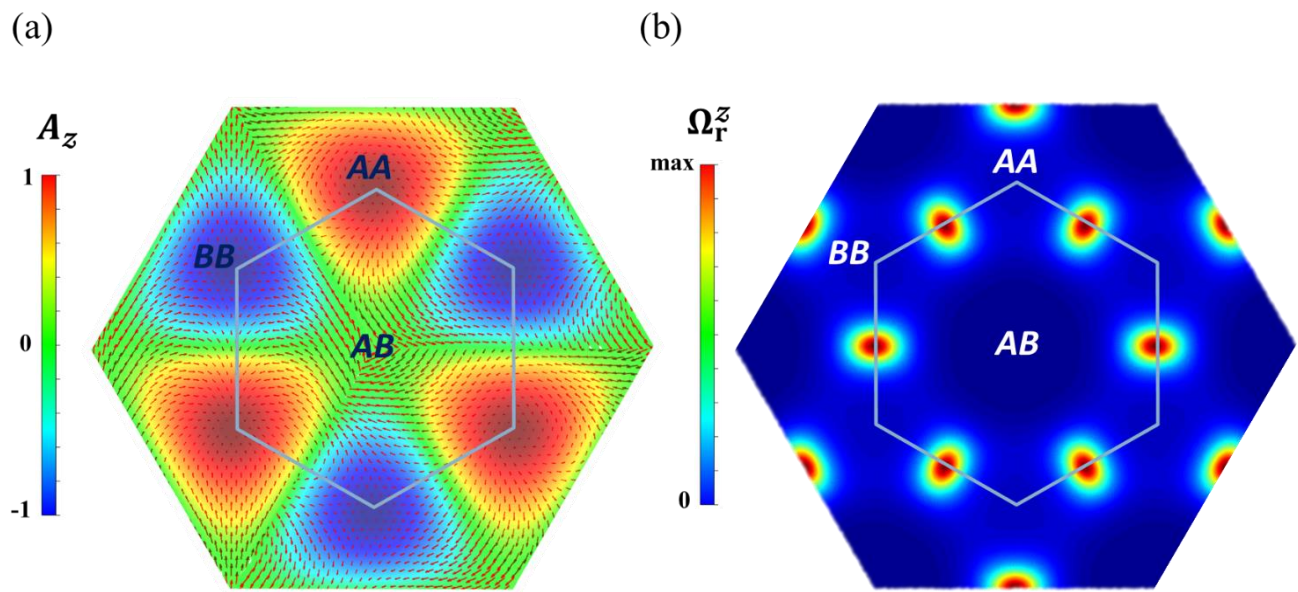

**Appendix**

The generic low-energy Hamiltonian of the bilayer system with **K**-valley can be written as

$$H_{bi} = \sum_{l=1,2} H_0^l + H_{int}, \tag{A1}$$

where $H_0^l$ is the monolayer Hamiltonian and $H_{int}$ is the interaction between the two layers with $l$ being the layer index. As twisting is imposed, one can use continuum description to handle modified interaction terms, which can be further attributed to the moiré potential incorporated within intralayer terms [28, 39].

Following this strategy, with constructed Wannier states we rebuild up Eq. (A1) as an effective tight-binding model [36] upon which we decompose the moiré potential as modified hopping energy multiplied by a phase modulation factor: $t_{mn} e^{i\phi_{mn}^l}$. The real-space atomic sites we configure can be an enlarged triangular lattice in which the lattice constant is comparable to interlayer distance. A simple consideration is that the phase modulation $e^{i\phi_{mn}^l}$ is induced by the intralayer hopping mediated by interlayer hopping. Then we have an effective tight-binding physics to simplify Eq. (A1) as [36]:

$$H_{eff} = \sum_{l=1,2} H_{eff}^l + 2H_{eff}^0 \tag{A2}$$

where $H_{eff}^l = \sum_{\langle m,n \rangle} t_{mn} e^{i\phi_{mn}^l} c_m^+ c_n^-$ and $H_{eff}^0$ can be described as the term independent of phase modulation. Here, $c_m^+(c_m^-)$ denotes the electron-creation (annihilation) operator centered at a triangular site $m$. $\langle m,n \rangle$ denotes the nearest-neighbor (NN) sites of the enlarged lattice. Due to the stacking variation given by relative rotation between layers, there should satisfies $\phi_{mn}^1 = -\phi_{mn}^2$, which give rise to the conjugation relations. One can optimize the parameters $t_{mn}$ and $\phi_{mn}$ to reproduce the band structures obtained from the continuum description.

Next, we consider a gapless case of ferrovalley system for simplicity [37]. Note that the degree of the freedom of a massless Dirac electron equals four, that is, two valley spin indices ($\tau$) and particle-hole indices ($\sigma$). In our case, an effective hoppling energy $t$ between NN-sites and an on-site Columbo repulsion energy $U$ are both introduced in an enlarged triangular lattice. In this way, for the low-energy physics of fourfold massless Dirac states, it is governed by a SU(4) model, where the Hilbert space is spanned by four indices. At each site, we can define the valley spin operator $\mathbf{T} = \frac{1}{2} c_{\tau\sigma}^+ \tau_{\tau\tau'} c_{\tau'\sigma}^-$ and due to particle-hole symmetry for Dirac electron, we also define the 'charge isospin' operator as $\mathbf{\Sigma} = \frac{1}{2} c_{\tau\sigma}^+ \sigma_{\sigma\sigma'} c_{\tau\sigma'}^-$. Here, $\tau$ and $\sigma$ are Pauli matrices respectively defined in valley and particle-hole subspace.

We start by writing the low-energy effective Hamiltonian into four fermion interaction form as below [40, 41]:

$$H_{eff}=J_{eff}\sum_{\langle m,n\rangle}\sum_{\tau,\tau',\sigma,\sigma'}(c^+_{m\tau\sigma}c^+_{n\tau'\sigma'}c^-_{m\tau'\sigma'}c^-_{n\tau\sigma}).\tag{A3}$$

Here $c^{+(-)}_{\tau\sigma}$ denotes the electron-creation (annihilation) operator in $\sigma$ and $\tau$ subspace and $J_{eff}$ describes the associated interaction energy between two NN sites, which comes from the net effect of the two competitive interactions. To be specific, the strong electron-electron correlation tends to align the valley spin of each site antiferromagnetically from the standard super-exchange: $H_{AFM}=\frac{t^2}{U}\sum_{\langle m,n\rangle}\sum_{\tau,\sigma}c^+_{m\tau\sigma}c^-_{m\tau'\sigma'}c^+_{n\tau'\sigma'}c^-_{n\tau\sigma}$, which is obtained by using standard $t/U$ expansion to order $t^2/U$ since $t\ll U$, where $U$ is the on-site repulsion energy. Nevertheless, a positive inter-site exchange (Hund's coupling) will keep the system as ferromagnetic state for the valley spin: $H_{FM}=J\sum_{\langle m,n\rangle}\sum_{\tau,\sigma}c^+_{m\tau\sigma}c^+_{n\tau'\sigma'}c^-_{m\tau'\sigma'}c^-_{n\tau\sigma}$ ($J>0$). Since both the super-exchange and the exchange interaction can be mapped into a Kugel-Khomskii type Hamiltonian [42] with exact SU(4) symmetry, we can express an effective exchange coupling as $J_{eff}=J-\frac{t^2}{U}$.

Through labeling tensor operator $\tau\otimes\sigma=c^+_{\tau\sigma}\tau_{\tau\tau'}\sigma_{\sigma\sigma'}c^-_{\tau'\sigma'}$ with Einstein summation convention, we map above Eq. (A3) into following equations:

$$H_{eff}=-J_{eff}\sum_{\langle m,n\rangle}\left(\frac{1}{2}+2\boldsymbol{\sigma}_m\cdot\boldsymbol{\sigma}_n\right)\left[\frac{1}{2}+2\tau^z_m\tau^z_n+(-\tau^x_m+i\tau^y_m)(-\tau^x_n-i\tau^y_n)+(\tau^x_m+i\tau^y_m)(\tau^x_n-i\tau^y_n)\right]\tag{A4}$$

In above, we find the two terms $(-\tau^x_m+i\tau^y_m)(-\tau^x_n-i\tau^y_n)$ and $(\tau^x_m+i\tau^y_m)(\tau^x_n-i\tau^y_n)$ in the brackets gives a cross-product as $\pm i(\boldsymbol{\tau}_m\times\boldsymbol{\tau}_n)_z$ when expanded. The phase modulation as mentioned comes from this purely imaginary term. This modulation with long period between NN-sites is so small ($\tau$ canting is weak) that the cross-product $\pm i(\boldsymbol{\tau}_m\times\boldsymbol{\tau}_n)_z$ can be attributed as $e^{i\phi_{mn}}$ with $\phi_{mn}=\pm\frac{(\boldsymbol{\tau}_m\times\boldsymbol{\tau}_n)_z}{\boldsymbol{\tau}_m\cdot\boldsymbol{\tau}_n}$. Here, the signs $\pm$ indicates that these two terms are indeed locked with top and bottom layers, which should be consistent with Eq. (A2).

A nonzero valley-contrasting phase $\varphi$ inherited in electron hopping will break SU(4) symmetry [35]. Now the antiferromagnetic term become $H_{AFM}=\frac{t^2}{U}\sum_{\langle m,n\rangle}\sum_{\tau,\sigma}e^{i(\varphi_{mn\tau}-\varphi_{mn\tau'})}c^+_{m\tau\sigma}c^-_{m\tau'\sigma'}c^+_{n\tau'\sigma'}c^-_{n\tau\sigma}+H.c.$ For $\tau=\pm$ we have $\varphi_{mn+}=-\varphi_{mn-}=\varphi$. The two terms in the brackets of Eq. (A4) become $e^{2i\varphi}(-\tau^x_m+i\tau^y_m)(-\tau^x_n-i\tau^y_n)$ and $e^{-2i\varphi}(\tau^x_m+i\tau^y_m)(\tau^x_n-i\tau^y_n)$ but the coefficient is $\frac{t^2}{U}$. Consequently, we have:

$$H_{eff}=-J_{eff}\sum_{\langle m,n\rangle}\left(\frac{1}{2}+2\boldsymbol{\sigma}_m\cdot\boldsymbol{\sigma}_n\right)\left(\frac{1}{2}+2\boldsymbol{\tau}_m\cdot\boldsymbol{\tau}_n\right)+\frac{2t^2}{U}\sum_{\langle m,n\rangle}\left(\frac{1}{2}+2\boldsymbol{\sigma}_m\cdot\boldsymbol{\sigma}_n\right)\left[\sin2\varphi(\tau^x_m\tau^y_n-\tau^y_m\tau^x_n)+(\cos2\varphi-1)(\tau^x_m\tau^x_n+\tau^y_m\tau^y_n)\right]\tag{A5}$$

Here, the second term breaks SU(4) symmetry into $SU(2)_L\times SU(2)_R\times U(1)$ symmetry.